%% file: CSLreply.tex
\documentclass[twocolumn,prl,aps,superscriptaddress,nofootinbib]{revtex4}

\usepackage{amsmath}
\usepackage{bm}
\usepackage{graphicx}
\usepackage{hyperref}
\usepackage{xspace}
\usepackage{amsfonts}

\makeatletter
\gdef\@fpheader{}
\g@addto@macro\bfseries{\boldmath}
\makeatother

\hypersetup{
    bookmarks=true,         
    unicode=false,          
    pdftoolbar=true,        
    pdfmenubar=true,        
    pdffitwindow=false,     
    pdfstartview={FitH},    
    pdfnewwindow=true,      
    colorlinks=true,        
    linkcolor=red,          
    citecolor=cyan,         
    filecolor=magenta,      
    urlcolor=blue,         
    linktocpage=true
}

\input{newcommands}

\begin{document}


\title{A response to criticisms on ``CMB Constraints Cast a Shadow on CSL Model''}

\author{J\'er\^ ome Martin} \email{jmartin@iap.fr}
\affiliation{Institut d'Astrophysique de Paris, UMR 7095-CNRS,
  Universit\'e Pierre et Marie Curie, 98bis boulevard Arago, 75014
  Paris, France}

\author{Vincent Vennin} \email{vincent.vennin@apc.in2p3.fr}
\affiliation{Laboratoire Astroparticule et Cosmologie, Universit\'e
  Denis Diderot Paris 7, 75013 Paris, France} \affiliation{Institut
  d'Astrophysique de Paris, UMR 7095-CNRS, Universit\'e Pierre et
  Marie Curie, 98bis boulevard Arago, 75014 Paris, France}

\date{\today}

\begin{abstract}
Our recent letter ``{\it Cosmic Microwave Background Constraints Cast
  a Shadow On Continuous Spontaneous Localization Models}''~\cite{Martin:2019jye} has recently been criticised in \Refa{Bengochea:2020efe} (see also \Refa{Bengochea:2020qsd}). In this reply, we explain
why the arguments presented in those articles are either incorrect or
a confirmation of the robustness of our results.
\end{abstract}

\maketitle

\section{Foreword}

Everybody agrees that Quantum Mechanics has successfully passed an
amazing number of experimental tests, yet there is a broad range of
opinions as to whether its theoretical status can be regarded as
satisfactory and self-consistent. One possible approach to this state
of affairs is the attempt to build alternatives to Quantum Mechanics,
the prototypical example being collapse models~\cite{Ghirardi:1985mt,
  Diosi:1988uy, Ghirardi:1989cn}. These theories are interesting
because, regardless of one's opinion about Quantum Mechanics, they
make different predictions and can, therefore, be falsified. 
Various setups aiming at testing collapse models
have now been studied and a review of their observational status can
be found in \Refa{Bassi:2012bg}. So far, no deviation from the
predictions of Quantum Mechanics has been found.

However, all experiments to date have been designed and performed in
the lab and the main goal of our letter~\cite{Martin:2019jye} was to
argue that cosmology can also be a crucial arena to test the viability
of collapse models, especially the Continuous Spontaneous Localisation
(CSL) model~\cite{Ghirardi:1989cn}. 
In CSL, the amplitude of the additional, non-standard, terms controlling
the dynamics of the collapse is generically proportional to the mass
and/or energy density. Therefore, one expects the effect to be maximum
for systems characterised by very large energy densities $\rho$. The
system with the largest $\rho$ that is possible to experimentally
probe in Nature is the very early universe during the phase of cosmic
inflation. Indeed, during inflation, $\rho$ can be as large as
$\rho_\mathrm{inf}\sim 10^{80}\mathrm{g}\times \mathrm{cm}^{-3}$,
which makes the early universe an ideal playground to further test
CSL.

In more details, the possibility to derive meaningful constraints from
inflation is based on the following line of reasoning. According to
inflation, the Cosmic Microwave Background (CMB) temperature and
polarisation anisotropies, and more generally, all the large-scale
structures observed in our universe, are nothing but quantum
fluctuations of the gravitational and matter fields, amplified by
gravitational instability and stretched to cosmological distances by
cosmic expansion during inflation. This simple mechanism has a great
explanatory power as it allows us, for instance, to understand in
details the most recent, high-accuracy, cosmological
observations. During inflation (and subsequently), the behaviour of
those quantum fluctuations is controlled by the Schr\"odinger
equation. Any modification of this equation thus changes how those
fluctuations evolve, with the potential danger to deliver predictions
in contradiction with the cosmological measurements. Moreover, as
already emphasised, one may expect those modifications to be very
substantial since the energy density during inflation is so large. As
a consequence, this opens up the possibility to probe CSL in different
regimes than those tested in the lab, and to derive meaningful
constraints on this class of theories.

Obviously, a legitimate concern is that the Physics of the very early
universe is uncertain and rests on speculative considerations. As a
consequence, even if it were possible to derive meaningful constraints
on CSL, those would necessarily be based on strong assumptions
and this would, therefore, greatly reduce their relevance. Fortunately
however, inflation relies on well-controlled physical mechanisms and
the situation is not as bad as it might seem. Indeed, the two main
mechanisms inflation rests on are (i) the fact that pressure
gravitates, as implied by General Relativity, and we have good reasons
to believe it is true, see \Refa{Martin:2015dha}, and (ii) quantum
parametric amplification by a classical source, namely exactly the
same mechanism responsible for the Schwinger
effect~\cite{Schwinger:1951nm}, the dynamical Casimir
effect~\cite{Davies:1976hi} (which has been observed in the
lab~\cite{Wilson_2011}), \etc.  Furthermore, over the last decades, the physical conditions that prevailed in the
early universe have been constrained by various high-accuracy
measurements, making cosmology not that different from conventional
Physics in the lab.

In our letter~\cite{Martin:2019jye}, we have carried out the program
described above and derived constraints on the CSL theory. This study
has been recently criticised in \Refa{Bengochea:2020efe} (see also \Refa{Bengochea:2020qsd}) and, below, we answer those criticisms in
detail. It is worth pointing out that
\Refs{Bengochea:2020qsd,Bengochea:2020efe} do not claim that our
calculations are incorrect but rather depict the assumptions on which
they are based as being too restrictive, preventing us from drawing
meaningful conclusions about CSL. In the following, we explain why we
disagree with these deductions.

\section{Choice of the collapse operator}

A first concern expressed in \Refa{Bengochea:2020qsd} is that, although
the collapse operator is identified with the smeared mass density in
CSL, in a general-relativistic context, the energy density $\rho$
might have to be replaced by a quantity related to the stress-energy
tensor $T_{\mu\nu}$, such as $T_\mu^\mu$ or
$\sqrt{T_{\mu\nu}T^{\mu\nu}}$. In fact, these choices all lead to
operators whose matrix elements are of order ${\cal
  O}(\rho_\mathrm{inf})$ and, from the arguments presented in the
foreword, an immediate conclusion is that it is unlikely to modify the
main result, at least in absence of very specific cancellations; and,
indeed, it is not difficult to reproduce our calculation for such
collapse operators, and to simply realise that the result is
unchanged. 

Let us now show how this can be concretely carried out. For linear perturbations, the collapse operator (let
us call it $\hat{C}$ in general) can always be linearly expanded onto
the Mukhanov-Sasaki variable $\hat{v}$ and its conjugated momentum
$\hat{p}$.  In our letter~\cite{Martin:2019jye}, when the collapse
operator is the energy density $\rho$, we find that, in Fourier space,
this expansion is of the form
\begin{align}
\hat{C}_{\bm{k}} = &
\ee^{-\frac{k^2 r_\uc^2}{2 a^2}} \Mp^2 H^2\sqrt{\frac{\epsilon_1}{2}} \left\lbrace \left[x_1 + x_2 \epsilon_1
  \left(\frac{aH}{k}\right)^2\right]\frac{\hat{v}_{\bm{k}}}{a\Mp} \right. \nonumber
\\ &  \left.  +\left[x_3 + x_4 \epsilon_1
  \left(\frac{aH}{k}\right)^2\right]\frac{\hat{p}_{\bm{k}}}{a^2\Mp H}
\right\rbrace\, ,
\label{eq:collapse_operator:decomposition}
\end{align}
where $\Mp$ is the reduced Planck mass, $a$ the
Friedmann-Lema\^itre-Robertson-Walker (FLRW) scale factor, $H$ the
Hubble parameter, ${\bm k}$ the wavenumber of the Fourier mode
considered, $\epsilon_1$ the first slow-roll parameter and $r_\uc$ the
CSL localisation scale. The quantities $x_1$, $x_2$, $x_3$ and $x_4$
are numbers of order one that entirely specify the model. As stressed
out in our letter~\cite{Martin:2019jye}, these numbers depend on the
gauge in which they are defined.\footnote{This does not mean that the collapse operators considered here are not gauge invariant, but rather that they coincide with the density contrast in different gauges.} For instance, in the longitudinal
gauge, during inflation, at leading order in slow roll, we had found
$x_1=-8$, $x_2=6$, $x_3=2$ and $x_4=-6$. Using
standard techniques in cosmological perturbation theory, one can show
that exactly the same
decomposition~\eqref{eq:collapse_operator:decomposition} is obtained
for the collapse operators proposed in \Refa{Bengochea:2020qsd}, though
with different $x_i$ numbers: when $C=T_{\mu}^{\mu}$, one finds
$x_1=20$, $x_2=-24$, $x_3=4$ and $x_4=24$; while
when $C=\sqrt{T_{\mu\nu}T^{\mu\nu}}$, these numbers are simply
multiplied by $-1/2$.

This result is not specific to the longitudinal gauge. In the comoving
gauge for instance, while we had found $x_1=-8$, $x_2=x_4=0$ and
$x_3=2$ for $C=\rho$, one obtains $x_1=20$, $x_2=x_4=0$ and $x_3=4$
for $C=T_{\mu}^{\mu}$, and these numbers are simply multiplied by
$-1/2 $ when $C=\sqrt{T_{\mu\nu}T^{\mu\nu}}$. The same is also true in
the flat gauge, where $x_1=-2$, $x_2=x_4=0$ and $x_3=2$ when
$C=\rho$, while these numbers are simply multiplied by $2$ when
$C=T_{\mu}^{\mu}$, and by $-1$ when $C=\sqrt{T_{\mu\nu}T^{\mu\nu}}$.

As a consequence, all results obtained in our
letter~\cite{Martin:2019jye} are simply multiplied by prefactors of
order one when working with the alternative collapse operators
proposed in \Refa{Bengochea:2020qsd}. Since we had found that, for all
choices (but one) of the density contrast, the correction to the CMB
power spectrum is at least $50$ orders of magnitude too large, 
operators of the form advocated in \Refa{Bengochea:2020qsd} cannot compensate for this discrepancy. 
In fact, this simple exercise just confirms the robustness of our
result~\cite{Martin:2019jye}.

Let us stress again that this $50$-orders-of-magnitude difference is
ultimately related to the very high energy at which inflation
proceeds. One would require a new physical scale to absorb these $50$
orders of magnitude, or substantial modifications to the theory; but
it is clear that the solution cannot merely come from discussing how
an energy density can be extracted from the stress-energy tensor,
which can only account for order-one modifications.

Finally, let us point out that if the goal of this discussion was to
find a collapse operator that is not ruled out by cosmological
experiments, we have already identified one in our
letter~\cite{Martin:2019jye}, namely the energy density evaluated in
the comoving threading. When derived from a more fundamental theory, CSL
should thus come with a prescription for the density contrast, which
we find has to match that particular choice (all other possibilities
being ruled out). This is a non-trivial condition that any attempt to
embed CSL in the general relativistic context should satisfy, and it
may help to guide such attempts. This was our main conclusion, which
we reiterate.

\section{Localisation in field space}

A second concern expressed in \Refa{Bengochea:2020qsd} is the fact
that, while for quantum particles, the notion of ``localisation''
naturally applies to their physical positions (hence the smearing
procedure is performed in physical space in CSL, over a distance
$r_\uc$), in a field-theoretic context, it may also apply to the value
of the fields themselves, and a smearing procedure in field space may
also have to be carried out, say over a field-value ``distance''
$\Delta$.

We first notice that the collapse operator is not the physical
position \textit{per se} in CSL, but rather the mass density
operator. As a consequence, although for quantum particles, this
induces the localisation of physical positions indeed, for fields,
this also entails the localisation of the field values. Indeed, in our
letter~\cite{Martin:2019jye}, we explicitly compute the wavefunction
associated to each Fourier mode of the Mukhanov-Sasaki field,
$v_{\bm{k}}$, and we find that $\Psi(v_{\bm{k}})$ gets peaked as the
collapse proceeds. Since this occurs when the wavelength associated to
$\bm{k}$ crosses out $r_\uc$, this means that $r_\uc$, a physical
distance, is also associated to a localisation process for the field
value, so the two mechanisms are not distinct.

It is then worth pointing out that in \Refa{Bedingham:2016aus},
a relativistic version of CSL is proposed, see appendix B, where the
field-space smearing procedure is carried out through the Bel-Robinsor
tensor, which is constructed from the Weyl tensor, which itself
vanishes for FLRW metrics. Therefore, in the context of cosmology,
that smearing procedure would become trivial. More generally, still in
appendix B of \Refa{Bedingham:2016aus}, it is then shown that this
smearing procedure reduces to the standard formulation of CSL
anyway. In the context of FLRW cosmology, this boils down to
introducing the scale factor at the required places, which gives
exactly the equation we have been using.

Although this does not preclude the possibility to build other
relativistic versions of CSL where field-space localisation plays a
non-trivial role, our main argument remains: those would have to pass
the test of cosmology and beat the $50$ orders of magnitude, which is
a non-trivial requirement.

In passing, it is also argued in \Refa{Bengochea:2020qsd} that the
amplitude of the CSL terms could be taken as time-dependent, which would
lead to different constraints. Again, since a fully satisfactory
version of relativistic CSL is not yet available, one can speculate on
the various additional features it could have, 
but let us point out that in
\Refa{Bedingham:2016aus}, following \Refa{Okon:2013lsa}, it is
proposed that the amplitude of the CSL terms depend on the Weyl
tensor, which would indeed induce space-time dependence of the
corrective terms. However, as already stressed, the Weyl tensor
vanishes in FLRW so this would lead to a constant amplitude of the CSL
terms. In case this happens, it is also argued in
\Refa{Bengochea:2020qsd} (and stated again in \Refa{Bengochea:2020efe})
that one could assume the corrective terms to depend on the Ricci
scalar. In FLRW, this is nothing but the Hubble parameter $H$, which
happens to be quasi-constant during inflation, leading to no time
dependence again. In fact, because of the maximal symmetry de-Sitter
space-times enjoy, introducing dependence of the parameters of the
theory on geometrical quantities cannot lead to effective time
dependence of the couplings, so this argument does not seem to apply
to the present context either. Again, this demonstrates that
introducing ``reasonable'' modifications that would be capable of
substantially modify our result is not trivial, a fact that reinforces the
robustness of our conclusions. Obviously, in the subsequent
radiation era, $H$ does depend on time and so would the CSL terms (if taken to be Ricci-dependent), as
we have studied (although in a slightly different context) in
\Refa{Martin:2012pea}.

\section{Semi-Classical Gravity}

\begin{figure*}[t]
\begin{center}
\includegraphics[width=0.99\textwidth]{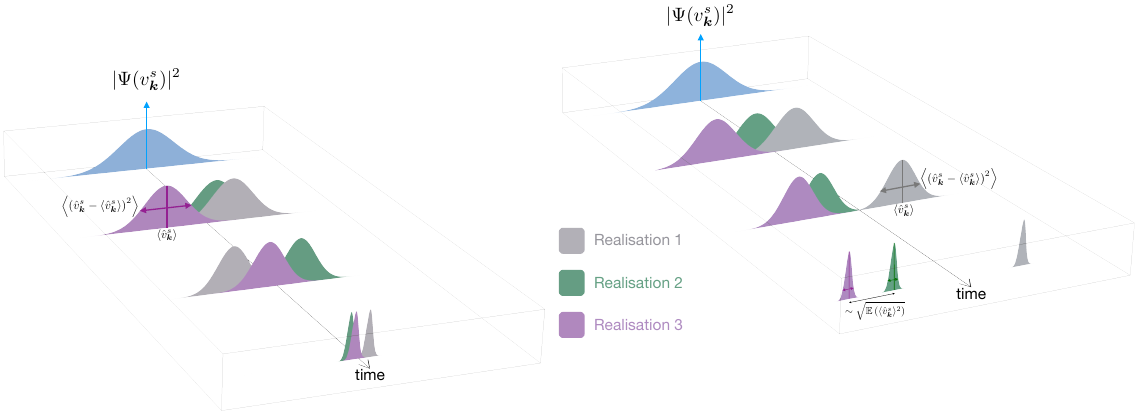}
\caption{In the framework of CSL, the wave-function of the
  perturbations $\Psi(v_{\bm k}^s)$ (where $s=\mathrm{R,I}$ denotes
  the real and imaginary parts of the Mukhanov-Sasaki variable $v_{\bm
    k}$), taken to be a Gaussian in the context of inflation, is a
  stochastic quantity. As a consequence, its quantum mean value
  $\langle \hat{v}_{\bm k}^s\rangle$ and quantum dispersion $\langle
  (\hat{v}_{\bm k}^s-\langle \hat{v}_{\bm k}^s\rangle)^2\rangle $ are random variables. In the present
  case, however, the quantum dispersion $\langle (\hat{v}_{\bm
    k}^s-\langle \hat{v}_{\bm k}^s\rangle )^2\rangle $ turns out to be
  a deterministic function. In the left panel, we have represented the
  stochastic ``trajectories'' of this wave-function for three
  different realisations. The means $\langle \hat{v}_{\bm k}^s\rangle
  $ evolve randomly while the dispersions continuously (and
  deterministically) decrease with time. At the final time, the
  dispersion of the means, $\mathbb{E}[\langle \hat{v}_{\bm
      k}^s\rangle^2]$ (the stochastic average of the means vanishes
  $\mathbb{E}[\langle \hat{v}_{\bm k}^s\rangle]=0$) is not small
  compared to the width of the wave-functions, $\mathbb{E}[\langle
    (\hat{v}_{\bm k}^s-\langle \hat{v}_{\bm
      k}^s\rangle)^2\rangle]=\langle (\hat{v}_{\bm k}^s-\langle
  \hat{v}_{\bm k}^s\rangle)^2\rangle$, and our criterion is not
  satisfied. In this case, the different wave-functions representing
  different realisations are not sufficiently separated to account for
  the emergence of different outcomes. In the right panel, on the
  contrary, our criterion is satisfied and different realisations do
  correspond to well-separated outcomes.}
\label{fig:collapse}
\end{center}
\end{figure*}

A last criticism put forward in
\Refs{Bengochea:2020qsd,Bengochea:2020efe}, which is not specifically
directed towards our works~\cite{Martin:2019jye,Martin:2019oqq} but
rather towards the whole community of primordial cosmologists and to
the standard formulation of inflation (which, admittedly, we use),
states that a quantisation of small fluctuations during inflation
cannot be carried out consistently and, instead, advocates the use of
semi-classical gravity based on the equation $G_{\mu \nu}=\langle
T_{\mu \nu}\rangle /\Mp^2$.

The question of whether gravity must and/or can be quantised is of
course a long-standing one. Although semi-classical gravity has
received many criticisms, the status of which are summarised \eg in
\Refa{Mattingly:2005nna} (there are even claims that it is already
ruled out either by actual, table-top, experiments such as Page and
Geilker experiment~\cite{Page:1981xe,Page:1981aj}, or that it leads to
superluminal signalling when combined with the standard collapse
postulate~\cite{Bahrami:2014gwa}, or that it is proven inconsistent by
thought experiments such as Eppley and Hannah's
experiment~\cite{Eppley1977-EPPTNO}), the modern consensus seems to be
that those arguments and experiments are not decisive enough to
invalidate semi-classical gravity. As a consequence, we agree that
arguing in favour of an inflationary mechanism based on this approach
might still be a defendable position even if it is held by a minority
of physicists.
However,
the criticism laid out in \Refs{Bengochea:2020qsd,Bengochea:2020efe}
against the standard approach of quantising the perturbations of both
the metric and the matter fields around a classical background, comes
with various statements that are worth commenting on.

Firstly, we notice that this criticism has nothing to do with CSL or
with how the collapse proceeds in the early universe: as a matter of
fact, effective collapse models have been used either in the context
of semi-classical gravity, see for instance \Refs{Perez:2005gh,
  Leon:2010fi}, or in the standard context, see
\Refa{DiezTejedor:2011jq}, thus showing that this issue is, in
some sense, disconnected from the main question discussed in our paper. On
general grounds, we think that, in order to investigate the
consequences of alternatives to a given standard formalism, it is
clearer to study one alteration at a time rather than to introduce several
variations at once.

\begin{figure}[t]
\begin{center}
\includegraphics[width=0.49\textwidth]{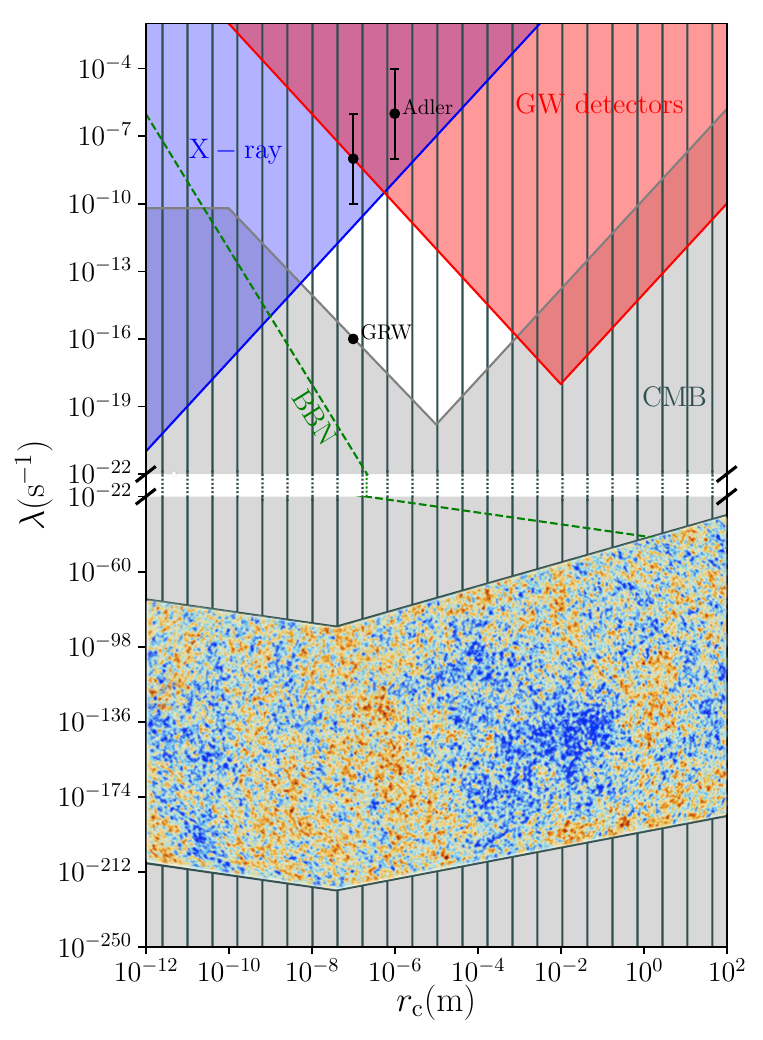}
\caption{Observational constraints on the two parameters $r_\uc$ and
  $\lambda$ of the CSL model. The white region is allowed by
  laboratory experiments while the unbarred region is allowed by CMB
  measurements (one uses $\Delta N=50$ for the pivot scale of the CMB,
  $H_{\mathrm{inf}}=10^{-5}\Mp$ and $\epsilon_1=0.005$). The two
  allowed regions are incompatible. The green dashed line stands for
  the upper bound on $\lambda$ if inflation proceeds at the Big-Bang
  Nucleosynthesis (BBN) scale. Taken from \Refa{Martin:2019jye}. }
\label{fig:mapCSL}
\end{center}
\end{figure}

Secondly, in our letter~\cite{Martin:2019jye}, we have introduced a
criterion for deciding whether or not the wave-function has collapsed,
which is based on the requirement that the average width of the
wave-function be much smaller than the dispersion of its mean value,
see~\Fig{fig:collapse} where we explain the rationale behind this
criterion. In \Refa{Bengochea:2020efe}, it is argued that such a
criterion may apply in the standard matter-metric quantisation
procedure, but is not the appropriate one in semi-classical
gravity. We do not really understand why this comment is relevant for
our work since we did not consider semi-classical gravity in our
letter. Furthermore, and more importantly, it is then stated that our
conclusions strongly rely on this criterion since, quoting
\Refa{Bengochea:2020efe}, it ``{\it has very relevant implications
  regarding what the values of the CSL parameters should be, and
  whether or not they are compatible with CMB observations}'' and
``{\it Their argument against CSL is that [...] it fails to achieve a
  sufficient localization of the relevant wave functions in the
  inflationary context}''. 
  At this point, there might be a 
misunderstanding of the calculation performed in our
letter~\cite{Martin:2019jye}, from which we reproduce Fig.~3, see
\Fig{fig:mapCSL} here. In this plot, we have represented the
constraints inferred from the CSL power spectrum, which is given by
\begin{align}
  {\cal P}_v(k)={\cal P}_v(k)\bigl \vert_\mathrm{std}
  \left\{1+\gamma \Delta {\cal P}_v(k)-\frac{
\mathbb{E}[\langle
    (\hat{v}_{\bm k}^s-\langle \hat{v}_{\bm
      k}^s\rangle)^2\rangle]}{\mathbb{E}[\langle \hat{v}_{\bm
          k}^s\rangle^2]}
\right\},
\end{align}
where ${\cal P}_v(k)\vert_\mathrm{std}$ is the standard, almost
scale-invariant, power spectrum, and $\gamma=8\pi^{3/2}r_\uc^3\lambda$, $\lambda$ being the mean rate of collapse.
We see that CSL introduces two types
of corrections, $\Delta {\cal P}_v(k)$ and $\mathbb{E}[\langle (\hat{v}_{\bm k}^s-\langle
  \hat{v}_{\bm k}^s\rangle)^2\rangle]/\mathbb{E}[\langle \hat{v}_{\bm
    k}^s\rangle^2]$, the explicit form of which is given in \Refa{Martin:2019jye}, and which turn out to be strongly scale dependent; the latter corresponding exactly to our
collapse criterion, as explained in~\Fig{fig:collapse}. We emphasise that this second type of corrections is
necessarily present in the CSL power spectrum regardless of the
interpretation it receives. If $\gamma \rightarrow 0$, the first
correction vanishes and the second one tends towards one since, then,
$\langle \hat{v}_{\bm k}^s\rangle \rightarrow 0$ (indeed, in absence
of CSL corrections, the dynamics is deterministic and the mean remains
zero). In this limit, the power spectrum vanishes, as expected since
only the CSL terms are able to break the homogeneity of the initial
vacuum state. In order to recover an almost scale-invariant power
spectrum the two corrections must be
sub-dominant. In~\Fig{fig:mapCSL}, the ``CMB-painted'' region
corresponds to a regime where CSL correctly accounts for the emergence
of primordial fluctuations, that is to say where the two types of
corrections are sub-dominant (hence, the power spectrum is almost
scale-invariant and the collapse criterion is satisfied). The region
dashed with vertical bars, on the contrary, represents a regime where
CSL fails to satisfactory describe the properties of cosmological
perturbations. The dashed region above the ``CMB-painted'' one
represents the region where the first correction dominates and the
second one is negligible. In this region, the power spectrum strongly
deviates from scale-invariance but the collapse criterion is
satisfied. The lower region dashed with vertical bars (that is the one
for $\lambda \lesssim 10^{-212}\mathrm{s}^{-1}$), below the
``CMB-painted'' one, corresponds to the opposite situation: the
correction $\Delta {\cal P}_v$ is negligible but the collapse
criterion is not satisfied. The values of the CSL parameters that are
in agreement with laboratory experiments, on the other hand, lie in
the white region. Now, the problem we highlight
in~\Refa{Martin:2019jye} is that this white area falls in the upper
dashed region, which corresponds to where the first type of CSL
corrections to the CMB power spectrum are too large but where the
second type of corrections are small. We see that the main conclusion
of our letter~\cite{Martin:2019jye} is in fact reached in a regime
where the collapse criterion is always satisfied and, hence, does not
have any discriminatory power. In other words, had we ignored our
collapse criterion, the incompatibility between the lab and CMB
constrains would have remained exactly the same. Therefore, claiming
that the collapse criterion (which, we still think, is well justified)
plays a role in ruling out CSL theories is clearly incorrect.

Thirdly, in \Refs{Bengochea:2020qsd,Bengochea:2020efe}, it is stated
that the standard approach where matter and metric are quantised is
problematic because field commutation relations are not compatible
with the full spacetime causal structure. We believe that this remark
refers to Eq.~(14.1.2) of Ref.~\cite{Wald:106274} where it is noted
that, in a theory of Quantum Gravity, one may expect the metric
quantum operator $\hat{g}_{\mu \nu}$ to satisfy $\left[\hat{g}_{\mu
    \nu}(x),\hat{g}_{\alpha \beta}(x')\right]=0$ when $x$ and $x'$ are
space-like separated, while the very definition of ``space-like
separated'' requires to specify the metric itself, rendering this
criterion ill-defined. Two important remarks are in order. First, let
us recall that what is merely done in the standard approach to
inflation is to quantise small fluctuations around a classical
background, as we do, for instance, for phonons on top of a classical
crystal. Therefore, none of the issues that usually plague attempts to
build theories of Quantum Gravity are present in this context (recall
that the energy scale of inflation is observationally known to be, at
least, five orders of magnitude below the Planck scale), and for small
linear fluctuations, the standard techniques of quantum field theory
can be applied safely. Therefore, we believe that this criticism does not
apply to the perturbative calculation of the inflationary power
spectrum. Second, in known constructions of Quantum Gravity, the
small-fluctuation limit precisely reduces to the standard approach
where fluctuations of both the metric and the matter fields are
quantised in a gauge-invariant way, and not to semi-classical gravity,
see \eg \Refa{Scherk:1974ca} in the case of string theory and
\Refa{Rovelli:2005yj} in the case of loop quantum gravity. As a
consequence, if one is looking for insight from Quantum Gravity, one
is naturally led to the standard matter-metric quantisation, and not
to semi-classical gravity.

Fourthly, in \Refa{Bengochea:2020efe}, another potential issue is put
forward, the so-called ``gauge problem'', which is related to the
problem of how the background and the perturbations are split. Let us
stress that this question has nothing to do with the quantisation of
perturbations and is already present at the classical level. As
noticed in \Refa{Bengochea:2020efe}, the Bardeen formalism offers an
elegant way to deal with this issue. \Refa{Bengochea:2020efe} argues
that it is still unsatisfactory because it is valid only at first
order in the perturbations. Even if it were correct, that would not be
a problem for calculations of the power spectrum, which do not go beyond that order.
But in any case, the gauge-invariant formalism for cosmological
perturbations has long been extended to higher orders, see \eg
\Refa{Malik:2008im} where relevant quantities are constructed in
several gauges at second order, and \Refa{Langlois:2005ii} where a
non-perturbative, covariant construction is derived. In fact, it seems
rather ironical that the issue of gauge invariance is brought by
\Refa{Bengochea:2020efe} into the debate, given that, contrary to the
standard approach, semi-classical gravity has a clear gauge
ambiguity. Indeed, since fluctuations in the metric and the matter
field are treated differently in semi-classical gravity, and because
gauge transformations mix those different types of fluctuations,
different gauges necessarily give rise to different
results~\cite{Markkanen:2014dba}. Therefore, the gauge burden seems to be rather on
the semi-classical gravity proponents.

\begin{figure}[t]
\begin{center}
\includegraphics[width=0.49\textwidth]{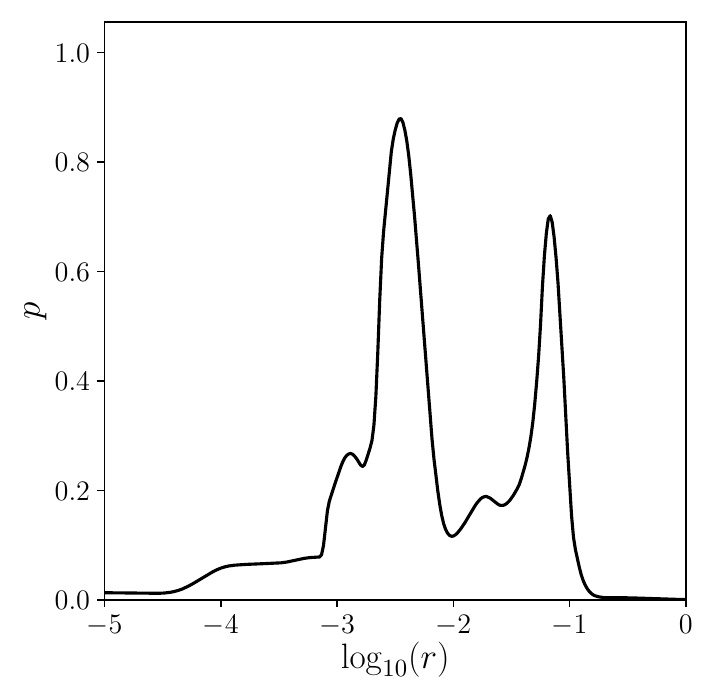}
\caption{Posterior distribution on the tensor-to-scalar ratio $r$,
  averaged over all physical single-field models of inflation as listed
  in \Refa{Martin:2013tda} (where each model is weighted by its
  Bayesian evidence), using the Planck 2013 data and the results of
  \Refa{Martin:2013nzq}.}
\label{fig:posterior:r}
\end{center}
\end{figure}

Finally, in \Refa{Bengochea:2020efe}, it is correctly noticed that one
way to observationally distinguish the two approaches would be to
measure the stochastic background originated from primordial quantum
gravitational waves. In the semi-classical approach, the signal is
indeed predicted to be so small that it should not be detected. If, on
the contrary, there is a detection, this would strongly support the
idea that small fluctuations must be quantised and would certainly
completely rule out the semi-classical approach. Primordial
gravitational waves can be observed either directly or by measuring
the B-mode polarisation in the CMB. For the moment, no signal has
been reported and, from the $2018$ Planck data
release~\cite{Akrami:2018odb}, we have for the tensor-to-scalar ratio
$r_{0.002}<0.1$ at $95\%$ Confidence Level (CL), an upper limit which
becomes $r_{0.002}<0.056$ if, in addition, the BICEP2/Keck Array BK15
data are used. Future experiments such as
LiteBIRD~\cite{Matsumura:2013aja} will be able to reach $r\sim
10^{-3}$. In \Refa{Bengochea:2020efe}, it is claimed that most
inflationary models predict values of $r$ that should already have
been seen, and that the standard treatment of inflation is already
under pressure, a conclusion that is clearly incorrect.  Such a
statement about what is predicted by ``most models'' would require to
actually count the number of models per value of $r$, something which
the authors of \Refa{Bengochea:2020efe} have not done. In order to
study this claim with well-justified methods, we display in
\Fig{fig:posterior:r} the posterior distribution on the
tensor-to-scalar ratio, obtained by averaging over all physical
single-field models of inflation as listed in {\it Encyclop\ae dia
  Inflationaris} (that contain $\sim200$ models),
see~\Refs{Martin:2013tda, Martin:2013nzq}, where each model is
weighted by its Bayesian evidence, obtained with the Planck 2013
data. This combines information about how the proposed models of
inflation populate different values for $r$, and observational
constraints (prior to the last Planck 2018 release and its combination
with BICEP2/Keck Array). One can see that there are roughly two
populations of models (two bumps in the posterior distribution): (i)
those predicting values of $r$ in the range $[10^{-2},10^{-1}]$ (those
were only weakly disfavoured in 2013 but they have been more strongly
discarded since then), and (ii) those predicting values of $r$ in the
range $[10^{-3},10^{-2}]$, which correspond to plateau models (a
prototypical example being the Starobinsky
model~\cite{Starobinsky:1980te}), which are not only in perfect
agreement with the data, but which will be probed by the next
generation of CMB experiments. The statement made in
\Refa{Bengochea:2020efe} that most models cannot account for the
current upper bound on $r$ is therefore ungrounded, as revealed by
this analysis of the landscape of inflationary models.

\section{Afterword}

Before closing this rebuttal, we would like to make a few additional
remarks. We, of course, agree with the authors of
\Refs{Bengochea:2020qsd,Bengochea:2020efe} that the correct
relativistic CSL theory is not yet known: we made this point very
clear in our letter~\cite{Martin:2019jye}. Therefore, exploring the
consequences of the CSL mechanism for cosmic inflation necessarily
involves some extrapolation. In fact, the whole discussion in
\Refa{Bengochea:2020efe} (see also \Refa{Bengochea:2020qsd}) boils down to the
question of what extrapolation is more likely, which, at this stage,
is subjective.
Facing this situation, we think it is more reasonable, at least in a
first step, to study the minimal extension and investigate what comes
out of it. Only if serious problems arise can we be forced to consider
more exotic possibilities.  Contrary to what is claimed in
\Refs{Bengochea:2020qsd,Bengochea:2020efe}, we do not think that
having a vast landscape of possibilities is an attractive feature of a
theory: instead, constrained theoretical frameworks lead to more
restrictive predictions, and can be better tested. Otherwise, the
Pandora's box is open and we loose any explanatory power. The
discussion presented here clearly shows that the results of
Ref.~\cite{Martin:2019jye} are robust and none of the suggestions
presented in Refs.~\cite{Bengochea:2020qsd,Bengochea:2020efe} seem
able to alter this conclusion.

Finally, 
the main point of our letter~\cite{Martin:2019jye} is not that, using
its most conservative extension, CMB data seem to cast a vague shadow
on CSL: it is rather that it does so by at least $50$ orders of
magnitude!  We agree that the formalism may have to be modified at high energies. Our main result is that these modifications
should have a drastic effect in order to overcome those $50$ orders of
magnitude. If we consider for instance, as we did in our
letter~\cite{Martin:2019jye}, a possible running of the fundamental
parameters, this running would have to be extremely strong. 
If we consider, instead, other
possible collapse operators, then, although we have shown that the
operators proposed in \Refa{Bengochea:2020qsd} are helpless, we already
had proposed a solution in our letter~\cite{Martin:2019jye}, and this
consists in considering the energy density evaluated in a very
specific threading (namely the ``comoving'' threading, leading to what is
called ``$\delta_\mathrm{m}$'' in \Refa{Martin:2019jye}).

In other words, the main result of \Refa{Martin:2019jye} is not at all
the claim that inflation rules out CSL (no such claim was ever made in
our letter) but rather that the corrections to the standard
framework that may appear at high energies must be very specific in
order to be compatible with cosmological data; this, of course, raises
the question of whether this is likely or even possible at all. On a
more positive note, this result can also be taken as a useful guide to
build extensions to the CSL framework. In any case, it is interesting
to see that cosmology can play a relevant role in developing our
understanding of Quantum Mechanics, a remark on which one should get
consensus from everyone.

\section{Acknowledgements}

It is a pleasure to thank Philippe Brax, David Langlois, Karim Noui, Patrick Peter, Christophe Ringeval,
Pierre Vanhove and David Wands for interesting discussions.

\bibliography{reference_reply}

\end{document}

%% file: newcommands.tex
\newcommand{\eg}{\textsl{e.g.~}}

\newcommand{\etc}{\textsl{etc.~}}

\newcommand{\ee}{e}

\newcommand{\sss}[1]{{\scriptscriptstyle{#1}}}

\newcommand{\uPl}{\mathrm{Pl}}

\newcommand{\uc}{\mathrm{c}}

\newcommand{\usssPl}{\sss{\uPl}}

\newcommand{\Mp}{M_\usssPl}

\newcommand{\beq}{\begin{equation}}
\newcommand{\eeq}{\end{equation}}
\newcommand{\bea}{\begin{eqnarray}}
\newcommand{\eea}{\end{eqnarray}}

\newlength{\wsingfig}
\newlength{\wdblefig}
\newlength{\wquadfig}
\newlength{\wtriplefig}
\newcommand{\Fig}[1]{Fig.~{\ref{#1}}}

\newcommand{\Refa}[1]{Ref.~{\cite{#1}}}
\newcommand{\Refs}[1]{Refs.~{\cite{#1}}}